# Slit-Induced Reflection Conversion Between Fundamental Lamb Modes in Elastic Plates with Low-Frequency and Broadband Response

Kaifei Feng and Pai Peng

School of Mathematics and Physics, China University of Geosciences, Wuhan 430074, China
*Corresponding author: paipeng@cug.edu.cn

## Abstract

We present a compact slit-based interface that enables reflection mode conversion in an elastic plate by transforming an incident lowest-order antisymmetric Lamb mode into a reflected lowest-order symmetric Lamb mode. The interface is realized by introducing inclined vacuum slits along the lateral sides of the plate. This geometry intentionally breaks symmetry and produces a strongly localized, vortex-like deformation near the slit tips, which provides an efficient coupling pathway from the antisymmetric incident response to a symmetric reflected component. The conversion is quantified in a consistent, field-based manner by extracting the reflected antisymmetric level from the standing-wave envelope on a probe line placed sufficiently far from the slits, and then inferring the converted symmetric energy under a lossless assumption with only two reflected propagating channels. The proposed design delivers strong low-frequency conversion and sustains broadband half-power operation, while parametric studies confirm that the performance is tunable and robust over a wide range of slit angles and widths.

Lamb waves in elastic plates provide an efficient platform for transporting mechanical energy and information over long distances with relatively low attenuation. Because plate-like components are ubiquitous in aerospace, marine, and civil structures, guided waves have become central to applications such as structural health monitoring, ultrasonic inspection, and vibration control. In these scenarios, it is often not sufficient to manipulate only the propagation direction; controlling the wave mode is equally important. The lowest-order antisymmetric mode (A0) and symmetric mode (S0) exhibit distinct displacement symmetries, dispersion characteristics, and energy partition between out-of-plane and in-plane motion, which directly affects how they interact with boundaries, discontinuities, and sensors. A reliable method for converting between A0 and S0 can therefore enable flexible signal routing, reduce modal crosstalk, and improve the detectability of damage signatures that couple preferentially to a specific mode.

However, strong and practically useful mode conversion remains challenging when low-frequency operation and broadband performance are required simultaneously. In the low-frequency regime, the wavelength is typically much larger than the characteristic size of engineered features, making scattering-induced mode mixing intrinsically weak. Moreover, the symmetry difference between A0 and S0 suppresses cross-coupling for many common perturbations, so conversion efficiencies often deteriorate as frequency decreases. On the other hand, approaches that can produce large conversion—such as those based on resonant elements or carefully tuned interference—tend to be narrowband and can become sensitive to geometric tolerances, which limits their robustness and practicality in real structures. These constraints motivate a mode-conversion strategy that is simultaneously effective at low frequency, broadband in operation, and simple to implement. Motivated by these limitations, we target a reflection-based converter that transforms an incident A0 Lamb wave into a reflected S0 wave using a compact structural modification that can be readily fabricated. Reflection conversion is particularly attractive because it can be realized locally without requiring long interaction lengths, and it can be integrated into plate edges or interfaces where guided waves are naturally manipulated. The central design challenge is to create a coupling region that intentionally breaks the symmetry that otherwise inhibits A0–S0 mixing, while sustaining this coupling over a wide frequency range rather than only at a single tuned point.

In this work, we develop a compact slit-based interface for A0-to-S0 reflection mode conversion in an elastic plate by introducing inclined vacuum slits along the lateral sides. The slits intentionally break symmetry and generate a strongly localized, vortex-like tip deformation that efficiently couples the incident antisymmetric response into a symmetric component radiating as the reflected S0 mode; the conversion is quantified by extracting the reflected A0 level from the far-field standing-wave envelope and inferring the converted S0 energy under a lossless two-channel reflection assumption. Normalized by the shear-wave speed and structural period, the interface delivers a near-unity conversion peak around $f/f_0 \approx 0.2$ and a broadband half-power operating band spanning approximately $f/f_0 = 0.2$–$0.5$ with CR about 0.5, while parameter maps of slit angle and width further confirm the tunability and robustness of both the low-frequency peak and the broadband band.

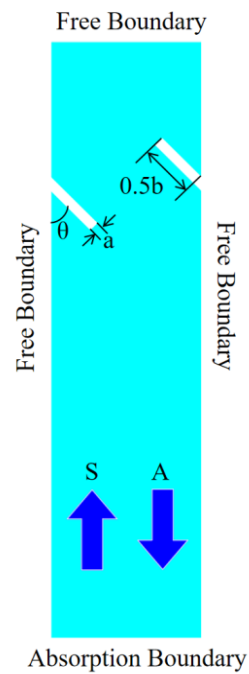

Fig 1. Computational model and geometry of the vacuum-slit interface in an elastic plate. Free boundaries are used on the top and side edges, and an absorbing boundary is applied at the bottom. The slits are defined by the width a, the length b (0.5b shown), and the inclination angle theta

measured from the y-axis; an A0 mode is incident from below and converts to a reflected S0 mode.

We consider guided-wave propagation in an isotropic, homogeneous elastic plate that is modified by introducing vacuum slits (void cuts) near the lateral edges to trigger reflection-induced mode conversion. The computational model follows the geometry sketched in Fig. 1. The top and the two side boundaries of the plate are traction-free, consistent with a free plate edge. An absorbing boundary is applied at the bottom of the domain to suppress spurious reflections from the truncation and to approximate a semi-infinite plate. Each slit is characterized by its full length b and width a, and its orientation is specified by the angle theta measured from the y-axis, where y denotes the primary propagation direction. The slits are voids, so their surfaces are also traction-free. A characteristic in-plane length p is used as the reference scale for reporting geometric parameters and for normalizing frequency. The plate material is modeled as lossless linear elasticity with density 1300, longitudinal wave speed 2540, and shear wave speed 1160.

The objective is to realize a controlled conversion from an incident A0 Lamb mode to a reflected S0 Lamb mode. In the simulations, an A0 mode is launched from the bottom and propagates upward toward the slit region. After interacting with the slits, the wave field below the structure consists of a superposition of an upward-traveling incident A0 component and downward-traveling reflected components. The key performance metric is the fraction of the incident A0 energy that is converted into the reflected S0 mode. Directly isolating the reflected S0 energy is not straightforward in the near field, so we infer it indirectly by first quantifying the reflected A0 component. Specifically, the incident and reflected A0 waves form a standing-wave pattern along a probe line placed sufficiently far below the slit region, where near-field effects have decayed and only propagating content remains. The amplitude modulation of this standing wave provides the reflection level of the A0 mode: a larger contrast between successive maxima and minima indicates a stronger A0 reflection. Under the assumptions of a lossless system and the absence of additional propagating modes in the frequency range of interest, any reduction in the reflected A0 energy relative to the incident A0 energy is attributed to conversion into the reflected S0 mode. This standing-wave-based procedure enables a robust and consistent extraction of the A0-to-S0 reflection conversion performance across frequency and geometry, while keeping the evaluation independent of the detailed near-field displacement complexity around the slits.

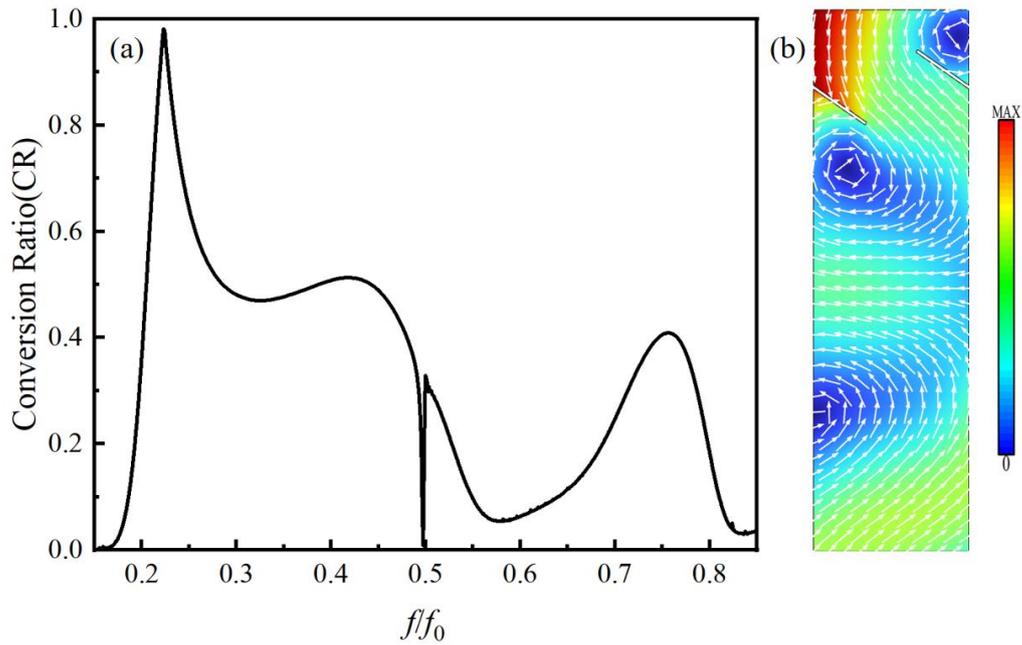

Fig 2. (a) Conversion ratio CR (incident A0 energy converted into reflected S0) versus normalized frequency $f/f_0$ ($f_0 = c_T/p$) for $a = 0.1p$, $b = 0.8p$, and theta = 45 deg. (b) Displacement field snapshot illustrating the localized tip motion responsible for the mode conversion.

Figure 2 demonstrates that the slit interface can redirect the incident A0 energy into the reflected S0 channel with two distinct operating regimes. First, a pronounced peak appears near the low normalized frequency of about 0.2, where the conversion ratio rises to nearly unity. This feature indicates that the structure can achieve almost complete mode conversion even when the wavelength is much larger than the characteristic geometric scale, which is precisely the regime where conventional scattering-based mode mixing is typically weak.

The physical origin of this low-frequency near-unity conversion is clarified by the excitation field shown in Fig. 2(b). The displacement concentrates strongly around the slit tips and forms a clear vortex-like localized motion. For a plate wave, such a localized rotational deformation is not a passive byproduct; it acts as an effective bending-moment and shear-couple source at the interface. Because the slits are inclined, the local deformation is not mirror-symmetric with respect to the propagation axis, so the scattered field cannot remain purely antisymmetric. In other words, the inclined void geometry breaks the symmetry that would otherwise keep A0 and S0 weakly coupled, and the localized vortex provides a compact region where bending and in-plane stretching are strongly interlinked. At the peak frequency, this coupling becomes particularly efficient: the structure preferentially channels the incident A0 energy into a symmetric radiating component, while the reflected A0 component is strongly suppressed. In the standing-wave-based evaluation, this manifests as a very small A0 reflection level and therefore an almost complete transfer of energy into the reflected S0 mode. Beyond the isolated peak, Fig. 2(a) also shows a broadband conversion band spanning roughly 0.2 to 0.5, where the conversion ratio stays around 0.5 with relatively modest variation. This sustained, moderate conversion level is consistent with a mechanism that is not purely a narrow resonance. The same geometric symmetry breaking and tip-localized deformation that enable strong coupling at the peak remain active over a wide frequency interval, continuously providing a pathway that mixes antisymmetric and symmetric responses upon reflection. In this range, the interface does not fully eliminate the reflected A0 component, but it maintains a persistent balance between the two reflected channels, so that approximately half of the incident A0 energy is systematically redirected into S0. Interpreted with the half-power, 3 dB criterion, this continuous interval therefore constitutes a broadband operating region for practical mode conversion. Finally, the sharp notch close to 0.5 indicates a frequency where the conversion pathway is temporarily quenched. A plausible interpretation, consistent with scattering from compact mode-coupling features, is that the symmetric radiation generated by the slit region undergoes a phase condition that drives destructive interference in the far field, effectively restoring an A0-dominant reflection for that specific frequency. The key point for the present study is that the notch sets the upper boundary of the main broadband band for the chosen geometry, while the overall response still exhibits both the low-frequency near-unity conversion and the wideband plateau enabled by symmetry breaking and vortex-like tip localization.

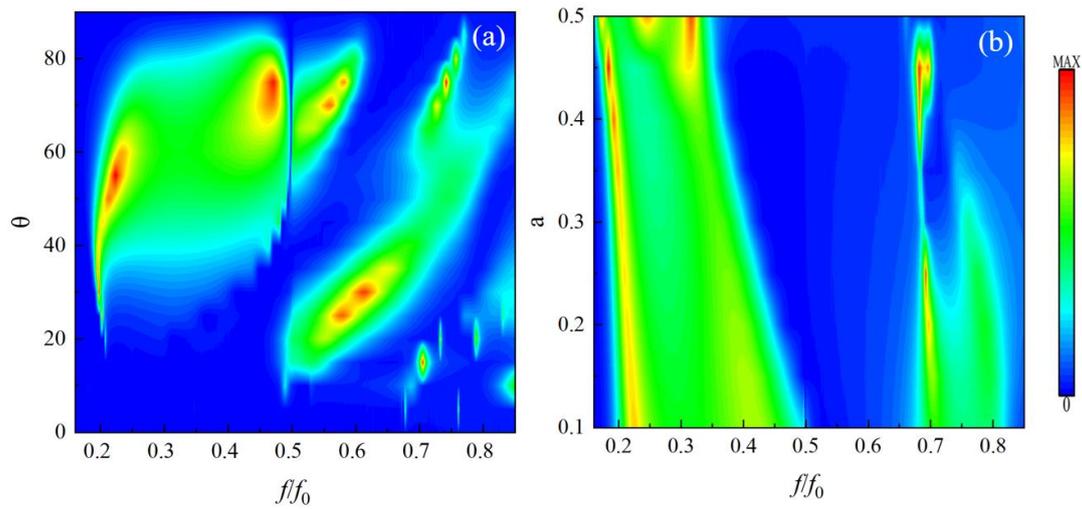

Figure 3. Parametric maps of the conversion ratio CR versus normalized frequency f/f0: (a) variation with slit angle theta; (b) variation with slit width a (other parameters fixed).

Figure 3 examines how the A0-to-S0 reflection conversion depends on two key geometric parameters: the slit inclination angle theta and the slit width a, while the remaining dimensions are kept the same as in the baseline design. The maps make two points clear. First, the strong low-frequency conversion observed near the normalized frequency around 0.2 is not an isolated accident of a single geometry; it persists over a broad range of theta and a. Second, the broadband plateau reported in Fig. 2 is also supported by an extended region in parameter space, which can be exploited to tune the operating band and improve robustness.

The angle map in Fig. 3(a) shows that efficient conversion requires a sufficiently oblique slit orientation. When theta is small, the slit is nearly aligned with the propagation direction, and the scattering remains closer to the original antisymmetric character, leading to weak coupling into the symmetric S0 channel. As theta increases into a moderate-to-large range, a wide high-response region emerges in the low-frequency regime, indicating that an oblique cut promotes stronger symmetry mixing upon reflection. Physically, increasing theta enhances the asymmetry of the local boundary perturbation experienced by the incident A0 wave and strengthens the localized rotational deformation around the slit tips. This tip-dominated vortex-like motion, already seen in Fig. 2(b), is precisely the type of localized deformation that can radiate a symmetric component and therefore feed the reflected S0 wave. The angle sweep also reveals that additional high-conversion islands appear at higher normalized frequencies, which suggests that, beyond the main low-frequency behavior, the slit region can support additional coupling conditions that depend on orientation.

The width map in Fig. 3(b) indicates that the low-frequency conversion peak near 0.2 is relatively insensitive to a, remaining pronounced across the scanned width range. In contrast, the broadband conversion region is more dependent on a. Smaller slit widths tend to maintain a broader interval of sustained, moderate conversion, consistent with a mechanism dominated by stress concentration and localized tip motion that remains effective over a wide frequency range. As the slit becomes wider, the conversion response becomes more structured and frequency-selective, with narrow high-conversion features emerging at higher frequencies. This trend is consistent with the idea that widening the void changes the local compliance and the phase relationship between scattered components, which can shift the balance between reflected A0 and converted S0 and, in some cases, confine strong conversion to narrower bands.

Taken together, Fig. 3 provides practical guidance for design selection. The baseline parameters used for Fig. 2 fall within a relatively broad high-response region, explaining why the structure exhibits both a near-unity low-frequency peak and an extended conversion plateau. At the same time, the maps indicate clear tuning knobs: theta primarily controls how strongly the interface breaks symmetry and activates the tip-localized rotational response, while a governs how broadband or frequency-selective the coupling becomes.

In conclusion, we demonstrated a compact slit-based interface that converts an incident A0 Lamb wave into a reflected S0 wave. The conversion is driven by symmetry breaking and strong tip-localized deformation, consistent with the vortex-like excitation field. The design achieves near-unity conversion at a low normalized frequency around 0.2 and maintains a broadband half-power conversion band from approximately 0.2 to 0.5 with CR about 0.5. Parameter sweeps further indicate that both the low-frequency peak and the broadband band are tunable and remain attainable over a wide range of slit angles and widths, providing practical design guidelines.